\pdfoutput=1
\documentclass[%
 aip,
 jcp,
 reprint,
 amsmath,amssymb,
]{revtex4-1}

\usepackage[utf8]{inputenc}
\usepackage[T1]{fontenc}
\usepackage{graphicx}
\usepackage{dcolumn}
\usepackage{bm}
\usepackage{mathptmx}
\usepackage{etoolbox}

\makeatletter
\def\@email#1#2{%
  \endgroup
  \patchcmd{\titleblock@produce}
    {\frontmatter@RRAPformat}
    {\frontmatter@RRAPformat{\produce@RRAP{*#1\href{mailto:#2}{#2}}}\frontmatter@RRAPformat}
    {}{}
}%
\makeatother

\begin{document}

\title{Derivation of the Sample-Size Scaling of TWO-NN Intrinsic-Dimension Estimates from Molecular Dynamics Trajectories}

\author{Riccardo Capelli}
\email{riccardo.capelli@unimi.it}
\affiliation{Department of Biosciences, Universit\`a degli Studi di Milano, Via Celoria 26, I-20133 Milan, Italy}

\date{\today}

\begin{abstract}
The intrinsic dimension of a dataset is the number of independent directions needed to describe the space occupied by its data. Nearest-neighbor estimators infer it from how the probability of finding a neighbor grows around each sampled point. Because the neighbor distances $r$ decrease as the sample grows, the estimated dimension can depend strongly on the number $N$ of available points, which in molecular dynamics is set by trajectory length and saving frequency. We derive the large-sample behavior of the TWO-NN estimator for data drawn from a smooth $d$-dimensional space. The typical nearest-neighbor distance scales as $N^{-1/d}$, and smooth deviations from local uniformity produce corrections in powers of $N^{-2/d}$. Isotropic noise adds a correction proportional to $N^{+2/d}$; for a five-dimensional hyperplane in 100 noisy features, the extended expansion recovers the dimension within a few percent when noise is moderate. We test the scaling on ten independent $100~\mu$s simulations of alanine dipeptide, represented by the pairwise distances among its ten heavy atoms, whose geometric dimension is $3n_{\mathrm{at}}-6=24$. Over the investigated range, the estimate does not depend systematically on the temporal spacing between configurations, but increases from $7.6$ to $15.6$ as $N$ grows from $10^2$ to $2\times10^5$. Extrapolations that retain corrections through $r^2$, $r^4$, and $r^6$ give limiting dimensions of $24.81$, $23.47$, and $25.30$, respectively. All three estimates lie within 6\% of the known dimension and bracket it, and their spread measures the uncertainty due to truncation. The derived scaling provides a practical route from finite-sample estimates to the geometric dimension.
\end{abstract}

\maketitle

\section{Introduction}

The intrinsic dimension (ID) of a dataset is the number of independent directions needed to describe the space occupied by its data. This number can be much smaller than the number of coordinates used to represent each point. Estimating ID is therefore useful in dimensionality reduction, manifold learning, and molecular simulation, where many Cartesian coordinates may describe a much smaller number of collective motions~\cite{Granata2016,Facco2017,Cazzaniga2026}. In molecular simulation, dimensionality estimates have been used to distinguish folded from unfolded states of a protein\cite{Allegra2020}, to characterize the free-energy landscapes of globular proteins\cite{Volkhardt2022}, and to analyze collective motions of macromolecules\cite{Cazzaniga2026}.

The dimension of an underlying continuous distribution cannot be read directly from a finite set of points. It must be inferred from how the points are arranged at a chosen distance scale. In practice, this scale is set by the number $N$ of sampled configurations, i.e., of data points. For a molecular dynamics trajectory, $N$ is fixed by the simulation length and by the saving frequency, which are practical choices rather than properties of the system. Nearest-neighbor methods set this scale automatically through the distances between nearby points~\cite{Levina2005,Facco2017}. TWO-NN, in particular, compares the distances from each point to its first and second nearest neighbors~\cite{Facco2017}. As the sample size increases, these neighbors become closer and the estimator probes finer details of the data distribution. The resulting ID can therefore change with sample size even when all points are drawn from the same distribution.

Previous work has used this dependence to study dimensionality across different spatial scales. At small $N$, only directions that vary over relatively large distances are resolved, whereas smaller-amplitude directions become visible as $N$ increases. Subsampling can therefore reveal scale-dependent dimensions and identify plateaus over which an effective dimension is observed~\cite{Facco2017,Denti2022}. These analyses establish sample size as a control of the spatial resolution probed by nearest-neighbor estimators. In generic data, the smallest scales are dominated by noise. The large-$N$ limit of the estimate then equals the number of noisy features, and the dimension of interest appears only at intermediate scales\cite{Facco2017,Denti2022}.

Here, we address a different question: whether the dependence of TWO-NN on $N$ can be assigned an explicit asymptotic form and used to infer the dimension reached at arbitrarily small distances. For points drawn from a smooth $d$-dimensional space, the typical distance between neighbors decreases as $N^{-1/d}$. We show that smooth corrections to the probability contained in a small neighborhood then generate a series in powers of $N^{-2/d}$. Similar powers appear in the finite sample bias of other nearest-neighbor quantities~\cite{Delattre2017}. We also show that isotropic noise adds a correction with the opposite sign in the exponent, so that the combined expansion can recover an intermediate dimension hidden by noise. The specific contribution of this work is therefore a functional relation between finite-sample TWO-NN estimates and their large-sample limit, rather than the observation that the estimated dimension can vary with scale. This has a direct practical consequence. A TWO-NN value is a property of the data at the scale probed by its $N$. Two values obtained from trajectories of different length or saving frequency cannot be compared directly, and a value obtained at a single $N$ should not be read as \emph{the} dimension of the system. Multiscale analyses remain the appropriate tool to identify dimensions that are stable over a range of scales; the present results describe the drift from which such a plateau must be distinguished.

We test this scaling with molecular dynamics (MD) trajectories of alanine dipeptide \textit{in vacuo}. Molecular data provide a useful benchmark because the dimension of the configuration space spanned by a complete distance representation is known independently. For $n_{\mathrm{at}}$ unconstrained atoms, pairwise distances remove global translations and rotations, leaving $3n_{\mathrm{at}}-6$ independent directions for a generic configuration~\cite{Asimow1978}. At any finite temperature, the Boltzmann distribution has nonzero density everywhere in configuration space, so this is the dimension approached at arbitrarily small scales. Lower values obtained at finite $N$ do not contradict it, because they count only the directions resolved at the corresponding scale. Our representation contains the distances among the ten heavy atoms and therefore has geometric dimension $24$. The simulations allow us to separate the effects of sample size and temporal spacing, and to test how reliably the known large-sample limit can be extrapolated from the available data.

\section{Large-sample scaling of TWO-NN}

Consider a point $x$ drawn from a smooth probability distribution on a $d$-dimensional space. Let $P_x(r)$ be the probability that another point lies within a distance $r$ from $x$. For a sufficiently small neighborhood,

\begin{equation}
\label{eq:local_mass}
    P_x(r)\simeq\rho(x)\omega_d r^d,
\end{equation}

where $\rho(x)$ is the probability density near $x$, $\omega_d$ is the volume of a unit ball in $d$ dimensions, and $r^d$ describes how the volume of the neighborhood grows with its radius. The exponent $d$ is therefore the local dimension.

Let $r_1(x)$ and $r_2(x)$ be the distances from $x$ to its first and second nearest neighbors, and define $\mu(x)=r_2(x)/r_1(x)$. If the density is approximately constant over these distances, the cumulative distribution $F(\mu)$ obeys~\cite{Facco2017}

\begin{equation*}
    1-F(\mu)=\mu^{-d}.
\end{equation*}

TWO-NN estimates $d$ from the slope of $-\log[1-F(\mu)]$ against $\log\mu$. The approximation becomes more accurate as the neighbors move closer to each sampled point.

For a sample containing $N$ independently drawn points (configurations), the typical nearest-neighbor distance $r_N$ can be estimated by requiring the expected number of points inside the neighborhood to be of order one,

\begin{equation*}
    NP_x(r_N)\sim1.
\end{equation*}

Using Eq.~(\ref{eq:local_mass}) gives

\begin{equation}
\label{eq:neighbor_scale}
    r_N\sim\left[\frac{1}{N\rho(x)\omega_d}\right]^{1/d}\propto N^{-1/d}.
\end{equation}

Thus, the distance scale probed by TWO-NN decreases in a predictable way as the sample grows.

At finite $r$, the density is not exactly constant and the underlying space need not be perfectly flat. These effects modify Eq.~(\ref{eq:local_mass}). Around an interior point of a smooth space, the first correction is quadratic in $r$,

\begin{equation}
\label{eq:mass_expansion}
    P_x(r)=\rho(x)\omega_d r^d\left[1+c_2(x)r^2+\mathcal{O}(r^4)\right].
\end{equation}

There is no term proportional to $r$. To see why, introduce a displacement $u$ from the center $x$. The contribution that is linear in $u$ changes sign between $u$ and $-u$, whereas the ball of integration contains both directions. The two contributions therefore cancel even when the density itself is not symmetric around $x$. This cancellation can fail near a boundary or a singular point.

The dimension measured at a finite radius can be expressed as the logarithmic slope

\begin{equation*}
    d_{\mathrm{eff}}(x,r):=\frac{\mathrm{d}\log P_x(r)}{\mathrm{d}\log r}.
\end{equation*}

Using Eq.~(\ref{eq:mass_expansion}) gives

\begin{equation*}
    d_{\mathrm{eff}}(x,r)=d+2c_2(x)r^2+\mathcal{O}(r^4).
\end{equation*}

TWO-NN does not calculate this derivative directly, but its neighbor ratios are generated by the same local probability $P_x(r)$ and inherit the same powers of $r$. Substituting the scale $r_N\propto N^{-1/d}$ therefore gives a large-sample expansion in powers of $N^{-2/d}$. After averaging over the sampled points, we write

\begin{equation}
\label{eq:finite_size_scaling}
\begin{aligned}
\mathrm{ID}_K(N)={}&\mathrm{ID}_{\infty}+\sum_{k=1}^{K}C_{2k}N^{-2k/\mathrm{ID}_{\infty}}\\
&+\mathcal{O}\left(N^{-2(K+1)/\mathrm{ID}_{\infty}}\right).
\end{aligned}
\end{equation}

Here, $\mathrm{ID}_{\infty}$ is the dimension approached as $N$ becomes large, $K$ is the number of correction terms retained, and the coefficients $C_{2k}$ collect the effects of local density variations, curvature, and the chosen distance measure. Their signs are not fixed. For a smooth distribution on a regular space, $\mathrm{ID}_{\infty}$ equals its geometric dimension.

Equation~(\ref{eq:finite_size_scaling}) is a large-sample expansion, not a general fitting function for arbitrary $N$. It applies only when nearest-neighbor distances are small enough for Eq.~(\ref{eq:mass_expansion}) to be accurate. Boundaries, singularities, disconnected components, or non-smooth noise can change the corrections and, in some cases, the limiting dimension itself.

The derivation above assumes that the data lie on a smooth $d$-dimensional space. In many applications, however, a few relevant degrees of freedom are embedded in a much larger space of features, all of which carry noise. The estimate then approaches the number of features at large $N$, and the dimension of interest appears only at intermediate scales. The same power counting can be extended to this idealized case.

Consider points on a smooth $d$-dimensional manifold embedded in a space of $D\gg d$ features, and add to each feature independent Gaussian noise with standard deviation $\sigma$. Let $r$ be the distance between two points without noise, as above, and $\tilde r$ the distance between the same two points with noise. If $\bm{\eta}$ denotes the difference between their noise vectors, each of its $D$ components has variance $2\sigma^2$, and
\begin{equation*}
\tilde r^2=r^2+2\,\mathbf{r}\cdot\bm{\eta}+|\bm{\eta}|^2.
\end{equation*}
The last term is a sum of $D$ squares: its mean is $2D\sigma^2$, and its relative fluctuations are $\sqrt{2/D}$. The cross term has zero mean and standard deviation $2\sqrt{2}\,\sigma r$. When $r\gg\sigma$ and $r^2\gg\sqrt{2/D}\,2D\sigma^2$, both fluctuations are negligible, and $\tilde r^2\simeq r^2+2D\sigma^2$.

Adding the same constant to all squared distances does not change the ranking of neighbors. The first and second neighbors are therefore the same with and without noise, and the distance without noise to the nearest neighbor still scales as $r_N\propto N^{-1/d}$. TWO-NN, however, measures $\tilde r$, so the local slope that it probes is
\begin{equation*}
d_{\mathrm{eff}}(x,\tilde r)=\frac{\mathrm{d}\log P_x}{\mathrm{d}\log\tilde r}=\frac{\mathrm{d}\log P_x}{\mathrm{d}\log r}\,\frac{\mathrm{d}\log r}{\mathrm{d}\log\tilde r}.
\end{equation*}
The first factor follows from Eq.~(\ref{eq:mass_expansion}) and equals $d+2c_2r^2+\dots$. Differentiating $\tilde r^2=r^2+2D\sigma^2$ gives $\tilde r\,\mathrm{d}\tilde r=r\,\mathrm{d}r$, so the second factor is $\tilde r^2/r^2=1+2D\sigma^2/r^2$. Hence
\begin{equation}
\label{eq:noise_slope}
\begin{aligned}
d_{\mathrm{eff}}(x,\tilde r)&=\left(d+2c_2r^2+\dots\right)\left(1+\frac{2D\sigma^2}{r^2}\right)\\
&=d+2c_2r^2+\frac{2dD\sigma^2}{r^2}+4c_2D\sigma^2+\dots
\end{aligned}
\end{equation}
Substituting $r_N\propto N^{-1/d}$, the second term gives the correction of Eq.~(\ref{eq:finite_size_scaling}), proportional to $N^{-2/d}$, whereas the third grows as $N^{+2/d}$. The last term is constant and shifts the intercept by an amount of second order in the ratio between the noise and the scale of density or curvature variations. In the range where both corrections are small,
\begin{equation}
\label{eq:two_sided}
\mathrm{ID}(N)\simeq d+C_2N^{-2/d}+B_2N^{+2/d}.
\end{equation}
The geometric correction dominates at small $N$ and the noise correction at large $N$. The dimension $d$ appears in between, and a fit that includes both terms can recover it.

The fluctuations neglected above generate higher positive powers of $N$. When $r_N$ becomes comparable with $\sigma$, or $r_N^2$ with $\sqrt{2/D}\,2D\sigma^2$, noise changes the ranking of neighbors and the estimate tends to the number of features. If boundaries contribute a term proportional to $r$, as for points uniform in a hypercube, the geometric correction starts at $N^{-1/d}$, and Eq.~(\ref{eq:noise_slope}) acquires a cross term proportional to $N^{+1/d}$.

The derivation also assumes independent samples from the equilibrium distribution, which consecutive MD configurations are not. At fixed $N$, changing the time interval $\Delta t$ between consecutive configurations changes their correlation, and the time window $N\Delta t$ covered by the sample, without changing the sample size. A systematic dependence on $\Delta t$ would therefore indicate that temporal correlations or incomplete sampling affect the estimate.

\section{Methods}

\subsection{Molecular dynamics}

Alanine dipeptide was simulated \textit{in vacuo} using the Amber99SB parametrization~\cite{Hornak2006} distributed with the PLUMED tutorials~\cite{Tribello2025}. Simulations were performed with GROMACS 2024.4~\cite{Abraham2015}. Electrostatic and van der Waals interactions were truncated at $1.2$~nm using a Verlet neighbor list, without long-range dispersion corrections. Bonds involving hydrogen atoms were constrained with LINCS using expansion order 4. Equations of motion were integrated with the leap-frog algorithm and a $2$~fs time step. The temperature was maintained at $300$~K with the velocity-rescaling thermostat~\cite{Bussi2007} and a coupling time of $0.1$~ps.

Ten independent replicas were initialized with different Maxwell-Boltzmann velocity distributions at $300$~K and propagated for $100~\mu$s each, yielding a total simulation time of $1$~ms.

\subsection{Configurational representation and geometric dimension}

Each configuration was represented by the $45$ pairwise distances among the ten heavy atoms. These distances do not change under a global translation or rotation, but they are not all independent. For a generic non-collinear configuration, the number of independent pairwise distances is~\cite{Asimow1978}

\begin{equation}
\label{eq:molecular_dimension}
    d_{\mathrm{geom}}=3n_{\mathrm{at}}-6-n_{\mathrm c},
\end{equation}

where $n_{\mathrm c}$ is the number of independent holonomic constraints among the atoms included in the representation. No bonds between the ten heavy atoms were constrained, so Eq.~(\ref{eq:molecular_dimension}) gives $d_{\mathrm{geom}}=24$. Constraints involving hydrogen atoms do not reduce the dimension because hydrogen atoms were not included in the representation. Equation~(\ref{eq:molecular_dimension}) is used here only as an independently known reference for the data-driven estimates. The representation contains positions only; momenta are not included, and all dimensions discussed here refer to configuration space. Perturbations of the Cartesian coordinates, including the 0.001~nm quantization of the compressed trajectories, still yield the distances of ten points in three dimensions. At scales above the quantization step, they therefore remain on the 24-dimensional manifold, so the large-$N$ limit is 24 rather than the number of features (45).

\subsection{TWO-NN estimation and sample-size analysis}

We denote by $N$ the number of configurations used in one estimate and by $\Delta t$ the time interval between two consecutive ones, a multiple of the 0.1~ns interval at which configurations were saved. For every pair $(N,\Delta t)$ and every replica, we used the configurations at times $0,\Delta t,\dots,(N-1)\Delta t$. The distances $r_1(x)$ and $r_2(x)$ from each configuration to its first and second nearest neighbors were calculated using the $45$ pairwise distances. The ratio $\mu(x)=r_2(x)/r_1(x)$ was then used to estimate ID. Following Ref.~\onlinecite{Facco2017}, the largest $10\%$ of the $\mu$ values were excluded from the linear fit because the upper tail of the distribution is sampled poorly. The slope was obtained by linear regression through the origin, with the empirical cumulative distribution $F_i=i/N$, $i=0,\dots,m-1$, as implemented in scikit-dimension 0.3.4\cite{Bac2021}; the current release uses $i=1,\dots,m$. For independent, Pareto-distributed $\mu$, this estimator has a statistical relative bias of $-1.45/N$, independent of $d$ (Supplementary Material, Sec.~II). The bias exceeds the statistical uncertainty for $N\le2\times10^3$, so all estimates were divided by $1-1.45/N$.

One TWO-NN estimate was obtained for each replica and each pair $(N,\Delta t)$. At fixed $N$, $A_N$ is the mean over the ten replicas of the estimates averaged over $\Delta t$, and its uncertainty is the standard error of these ten independent replica means. The resulting values of $A_N$ and their uncertainties were used in the sample-size fits.

Equation~(\ref{eq:finite_size_scaling}) was fitted with $\mathrm{ID}_{\infty}$ and all correction coefficients treated as free parameters. We considered $K=1$, $2$, and $3$, corresponding to corrections through $r^2$, $r^4$, and $r^6$. Estimates at different $N$ were obtained from the same trajectories and can share configurations, so they are correlated. These correlations were not included in the fits. A bootstrap over the replicas, which preserves them, gives intervals for $\mathrm{ID}_{\infty}$ close to those of the fits. The reported reduced chi-square values should therefore be read as descriptions of the residuals, rather than as formal statistical tests.

\section{Results and Discussion}

\subsection{Alanine dipeptide as a known-dimension benchmark}

\begin{figure}[h!]
  \includegraphics[width=\linewidth]{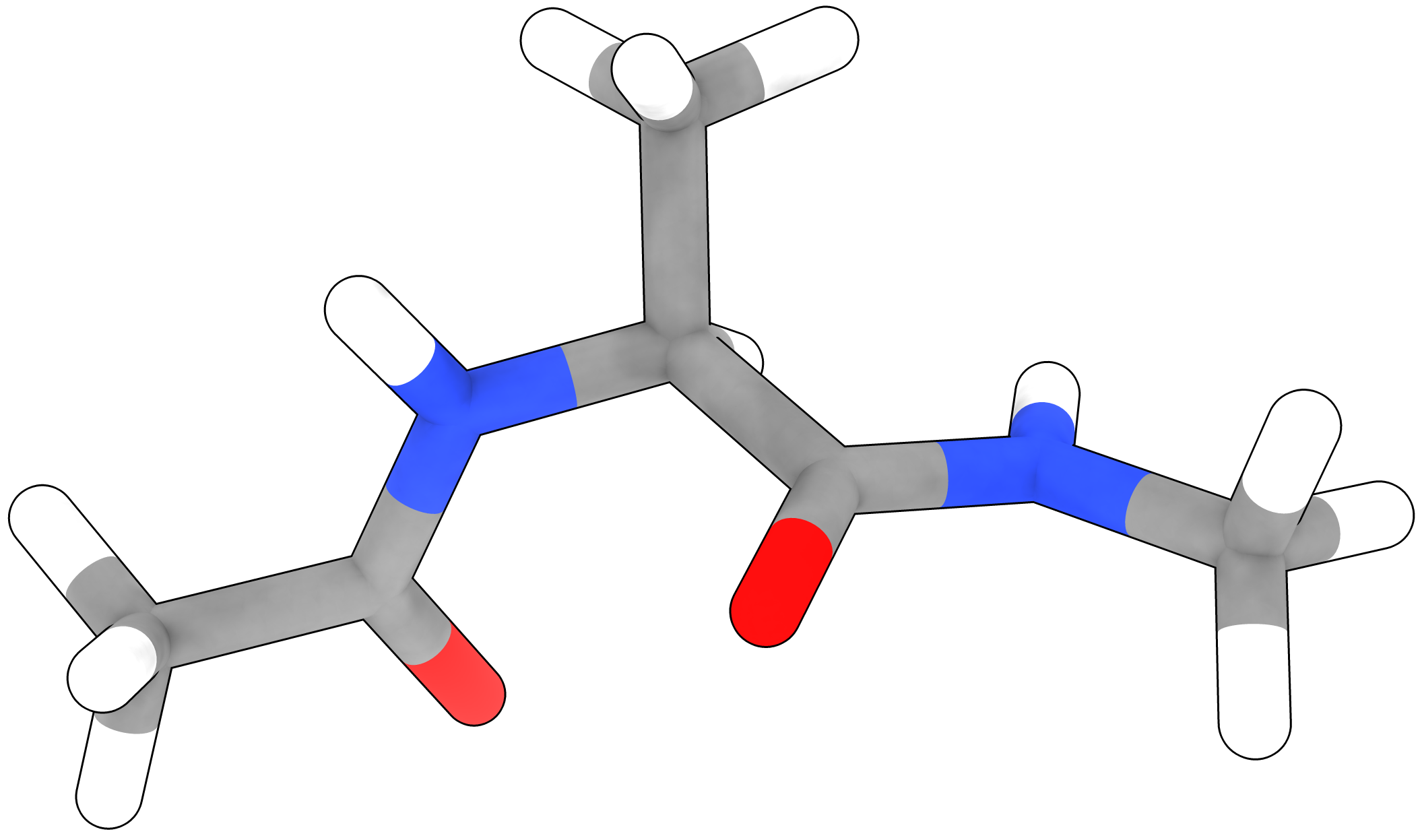}
  \caption{Alanine dipeptide (ACE--ALA--NME). Atoms are colored by element (gray: C, white: H, blue: N, red: O). The ten heavy atoms define the $45$ pairwise distances used for the ID estimate. Hydrogen atoms are shown but are not included in the representation.}
  \label{fig1}
\end{figure}

Alanine dipeptide provides a small but nontrivial system for which the dimension of the chosen representation is known. The combined unbiased trajectories sample both the $C7_{\mathrm{eq}}$ and $C7_{\mathrm{ax}}$ basins of the gas-phase molecule~\cite{Bolhuis2000} (see Figure S1). The ID analysis does not use these conformational labels or selected dihedral angles; it uses the complete set of pairwise distances among the ten heavy atoms shown in Fig.~\ref{fig1}.

\subsection{Temporal spacing has little effect over the investigated range}

\begin{figure}[h!]
  \includegraphics[width=\linewidth]{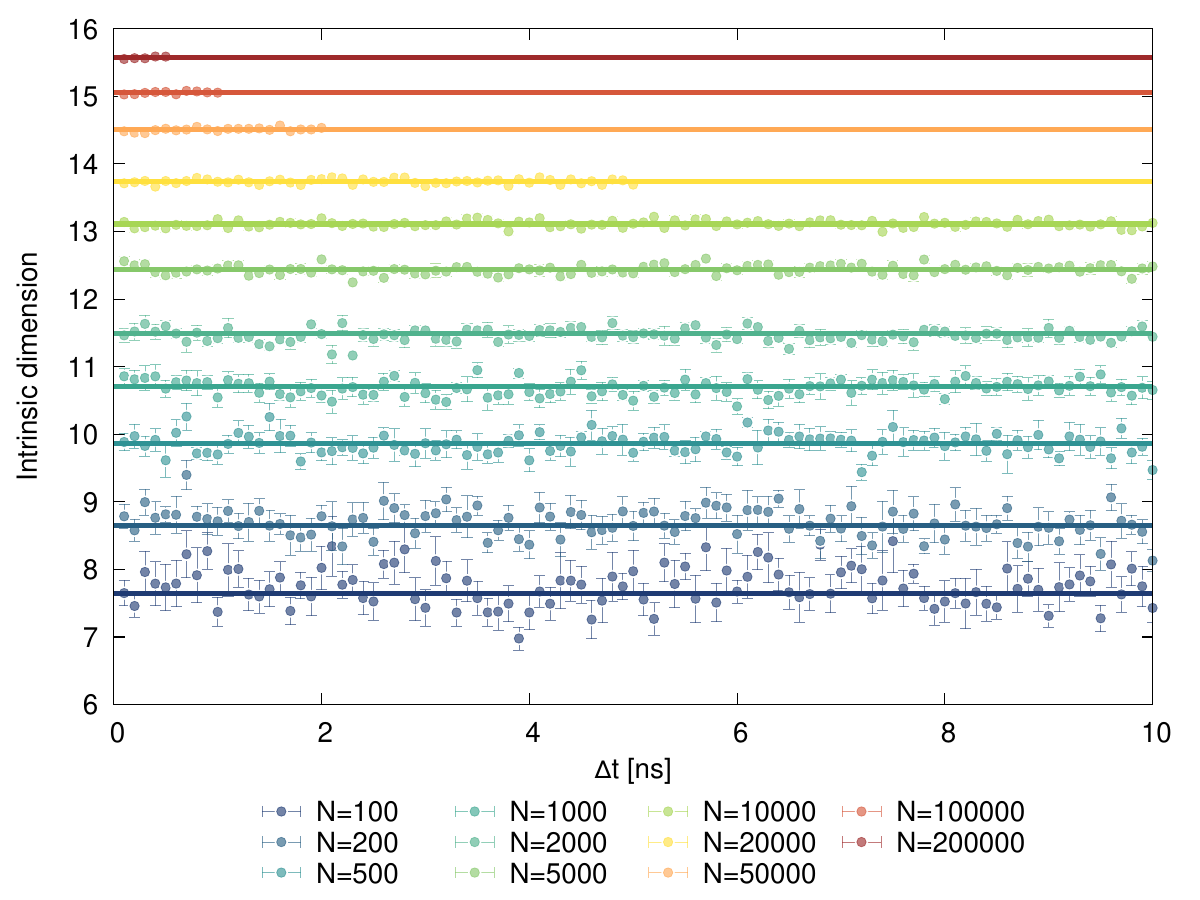}
  \caption{TWO-NN estimate as a function of the temporal spacing $\Delta t$ for different sample sizes $N$. Symbols show the estimates obtained from the ten independent replicas. All estimates are bias-corrected, and solid lines show the values $A_N$ defined in the Methods. Fewer time intervals are available at large $N$ because each trajectory has finite length.}
  \label{fig:id_vs_dt}
\end{figure}

Figure~\ref{fig:id_vs_dt} shows the TWO-NN estimate as a function of the time interval $\Delta t$ between consecutive configurations. Temporal correlations can affect dimension estimates based on nearest neighbors~\cite{Theiler1986}. Here, however, changing $\Delta t$ produces no systematic shift beyond the variation among replicas. This does not show that temporal correlations are unimportant in general. It shows that, for these trajectories and time intervals, their effect is smaller than the effect of sample size.

The weak dependence on $\Delta t$ is a property of the present system and trajectories. TWO-NN is local: each ratio $\mu$ depends only on the neighborhood of one configuration. A sample confined by kinetic barriers to part of configuration space is therefore estimated as the visited region alone. The estimate then describes only the visited states, whose local dimension may differ from that of the full distribution. Moreover, the $N$ points are spread over a smaller region, so the local density is higher and the sample behaves as a larger one; since the estimate increases with $N$, it is shifted upward. Both effects depend on the time window $N\Delta t$, and therefore appear as a dependence on $\Delta t$ at fixed $N$ until the window is long enough to explore the relevant states. In the present trajectories, alanine dipeptide leaves the $\phi<0$ region ($C7_{\mathrm{eq}}$ and C5) after 1.75~$\mu$s on average, and the $C7_{\mathrm{ax}}$ state accounts for 2.8\% of the population (Supplementary Material, Sec.~III). Since the windows $N\Delta t$ range from 10~ns to 100~$\mu$s, the shortest ones usually miss $C7_{\mathrm{ax}}$. Because this state carries less than 3\% of the population, a sample that misses it differs little from a complete one. For systems with slowly interconverting states of comparable population, a dependence on the window should be expected.

\subsection{The estimated dimension increases with sample size}

\begin{figure}[h!]
  \includegraphics[width=\linewidth]{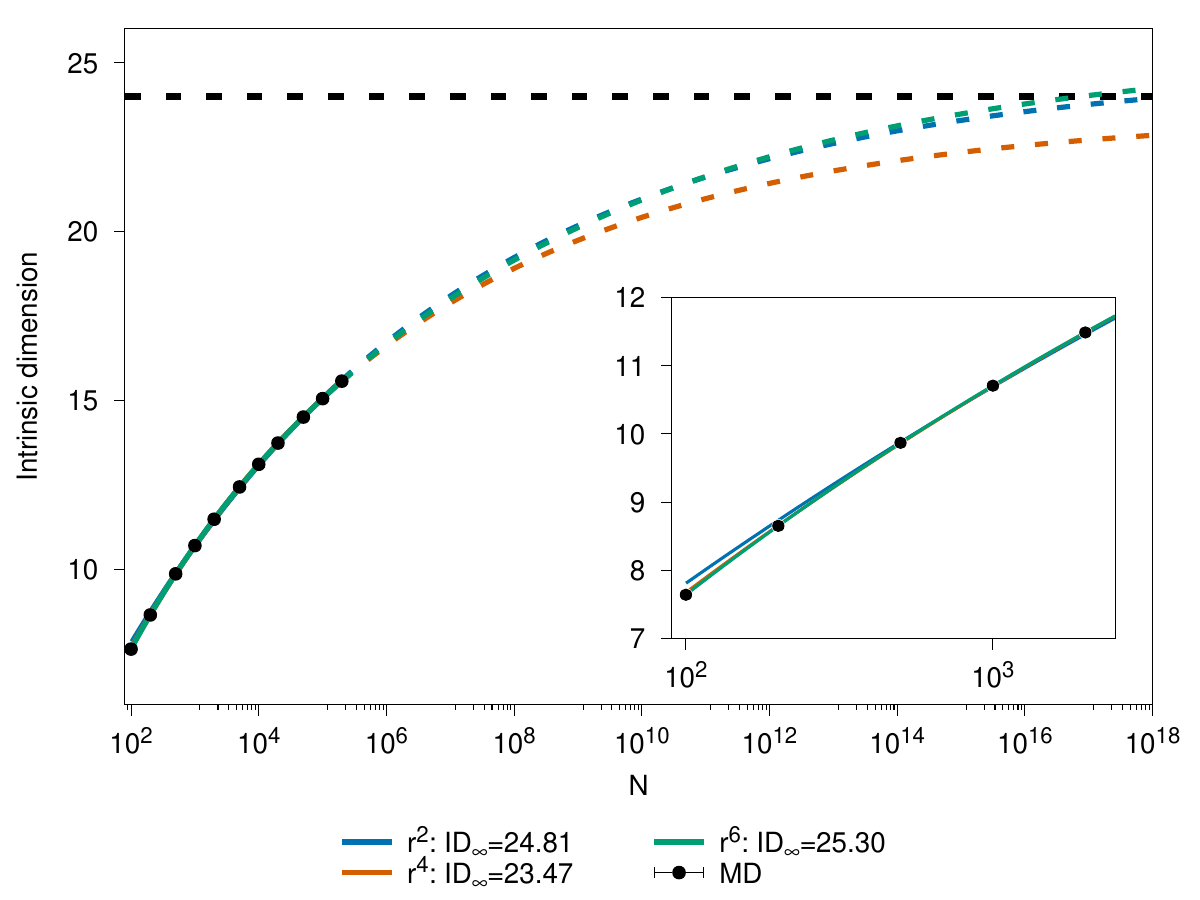}
  \caption{Dependence of the TWO-NN estimate on sample size. Each point is the value $A_N$ defined in the Methods; error bars are its standard error over the ten replicas. Curves show fits of Eq.~(\ref{eq:finite_size_scaling}) with corrections through $r^2$ ($K=1$, blue), $r^4$ ($K=2$, orange), and $r^6$ ($K=3$, green). Solid lines cover the measured range and dashed lines show the extrapolations. The inset enlarges the measured range and shows the systematic deviation of the $K=1$ model at small $N$. The horizontal dashed line marks the known geometric dimension $d_{\mathrm{geom}}=24$.}
  \label{fig:id_vs_N}
\end{figure}

The dependence on sample size is much stronger. The estimated dimension increases monotonically from $7.64$ at $N=100$ to $15.57$ at $N=2\times10^5$ (Fig.~\ref{fig:id_vs_N}). With more configurations, the nearest neighbors become closer and TWO-NN resolves molecular motions that are not visible at coarser resolution. The largest measured value remains well below the known geometric dimension $d_{\mathrm{geom}}=24$, and the data do not show a plateau. The simulations therefore do not reach the large-$N$ regime directly.

\subsection{The extrapolation locates the large-sample limit near the geometric dimension}

To infer the geometric dimension of the dataset we can perform an extrapolation using Eq.(\ref{eq:finite_size_scaling}). As expected, the extrapolated dimension changes when additional correction terms are included. The $K=1$ model gives $\mathrm{ID}_{\infty}=24.81\pm0.02$. The $K=2$ model gives $\mathrm{ID}_{\infty}=23.47\pm0.09$, whereas the $K=3$ model gives $\mathrm{ID}_{\infty}=25.30\pm0.74$ (Table S1). Despite the directly measured dimensions ranging only from $7.64$ to $15.57$, all three extrapolations lie within $6\%$ of the independently known value $24$ and collectively bracket it. This agreement indicates that the scaling identifies the correct geometric dimension from data that have not yet reached the corresponding plateau. We use the range $23.47$--$25.30$ as an estimate of the systematic uncertainty associated with the truncation. The $K=3$ value exceeds the maximum dimension of the representation, confirming that the individual extrapolations should be interpreted together rather than as separate exact determinations.

The spread among the extrapolations has a simple origin: the corrections decay very slowly. If the limiting dimension is $24$, the leading correction is proportional to $N^{-1/12}$. Increasing the sample size from $100$ to $200,000$, a factor of $2000$, reduces this term by only a factor of approximately two. Over such a limited change, a shift in $\mathrm{ID}_{\infty}$ can be offset by a change in the correction coefficients. Adding more terms improves the description of the measured points but also changes the extrapolated value, making the range across truncations a useful measure of model uncertainty.

The present benchmark therefore shows that the scaling can turn finite sample estimates far below the geometric dimension into a useful estimate of its large-sample range.

\subsection{Noise and intermediate scales}

\begin{figure}[h!]
  \includegraphics[width=\linewidth]{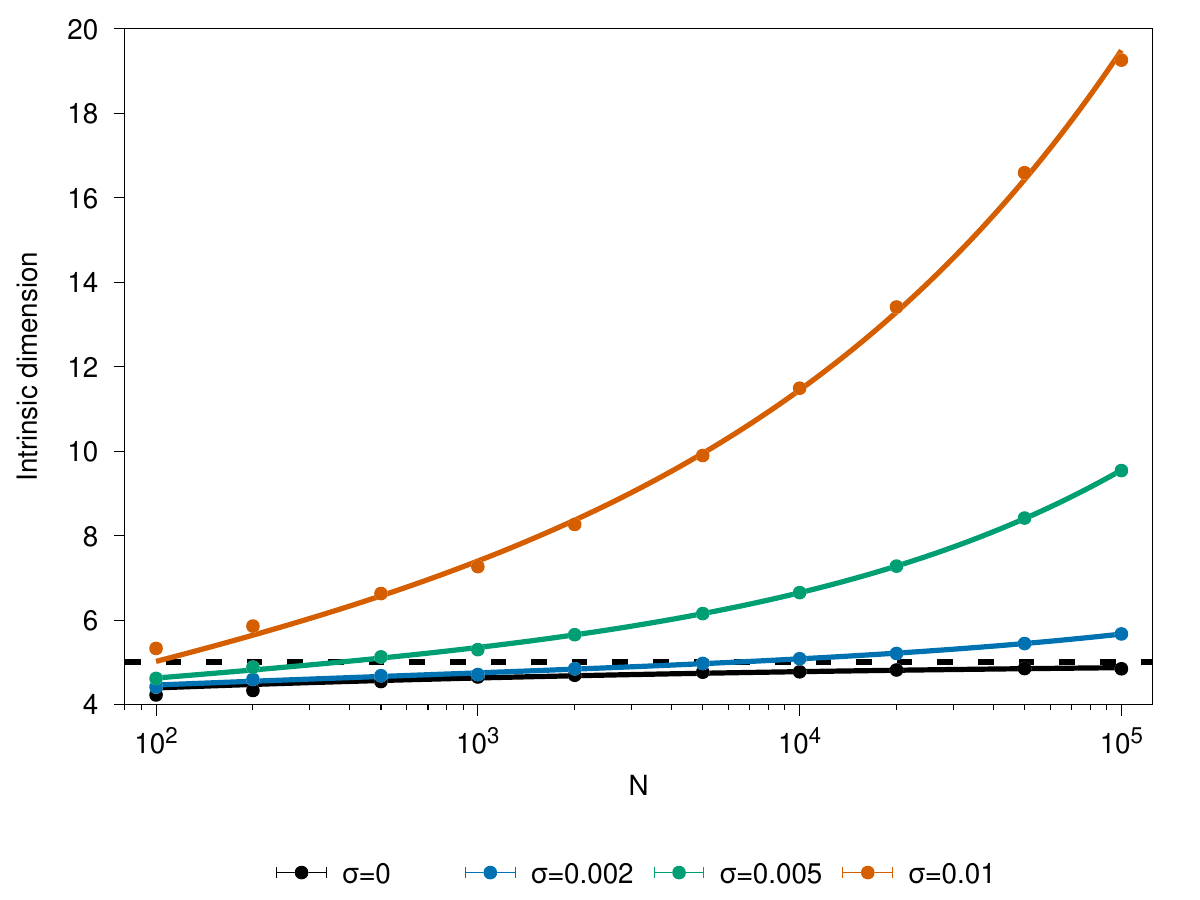}
  \caption{TWO-NN estimate (bias-corrected) for points uniform in a five-dimensional hypercube embedded in $D=100$ features, with Gaussian noise of standard deviation $\sigma$ on each feature. Lines: fits with the geometric correction only ($\sigma=0$) and with the geometric correction and the noise term of Eq.~(\ref{eq:two_sided}) ($\sigma>0$); the fitted dimensions are reported in Table~S2. The horizontal dashed line marks the true dimension, 5.}
  \label{fig:noise}
\end{figure}

To test whether the expansion can separate a genuine intermediate dimension from the drift induced by noise, we analyzed points uniform in a five-dimensional hypercube embedded in $D=100$ features, with Gaussian noise of standard deviation $\sigma$ added to all features (Fig.~\ref{fig:noise} and Table~S2). Because of the boundaries, the leading geometric correction is proportional to $N^{-1/5}$ (Sec.~II). Without noise, the estimate approaches 5 from below, and Eq.~(\ref{eq:finite_size_scaling}) with this correction recovers 5.0. With noise, the estimate grows steadily with $N$: for $\sigma=0.005$ it reaches 9.5 at $N=10^5$. A fit with the geometric correction only cannot describe these data: that correction makes the estimate level off at large $N$, whereas with noise the estimate grows increasingly fast. The fit returns $d=6.3$ and 40 for $\sigma=0.002$ and 0.005, and does not converge for $\sigma=0.01$ (Table~S2). Adding the noise term of Eq.~(\ref{eq:two_sided}) gives $4.98\pm0.04$ and $5.06\pm0.03$ for $\sigma=0.002$ and 0.005, and describes the data within their uncertainty. For $\sigma=0.01$, the conditions of Sec.~II are violated at large $N$: the sampled range contains no window between the noise scale and the scale of the geometric corrections, and the fit returns about 6 and does not describe the data. The expansion therefore recovers the intermediate dimension to within a few percent when the sampled scales include such a window, and the quality of the fit indicates when it is absent.

\section{Conclusion}

We derived the large-sample dependence of TWO-NN for a smooth $d$-dimensional distribution. The typical distance between neighbors decreases as $N^{-1/d}$, and smooth departures from a locally uniform distribution generate corrections in successive powers of $N^{-2/d}$. Isotropic noise adds a correction proportional to $N^{+2/d}$. For a noisy five-dimensional hyperplane, the resulting expansion recovered the intermediate dimension to within a few percent whenever the sampled scales included a window between the noise scale and the scale of the geometric corrections. Whereas subsampling analyses use changes with $N$ to identify dimensions resolved over different spatial scales, the present result assigns this dependence an explicit asymptotic form that connects finite-sample TWO-NN estimates to their geometric limit.

For alanine dipeptide, the estimate is insensitive to the investigated temporal spacings but rises from approximately $7.6$ to $15.6$ over the available range of $N$. Extrapolations with different numbers of correction terms give limiting dimensions between $23.47$ and $25.30$, bracketing the known geometric value of $24$. Thus, although the geometric plateau is not observed directly, the extrapolation locates the limiting dimension near the geometric value. The variation among truncation orders provides an estimate of its systematic uncertainty.

In practice, a TWO-NN estimate depends on how many configurations were used to compute it. In our alanine dipeptide dataset, the estimate is 13.1 with $10^4$ configurations and 15.1 with $10^5$, although the system is exactly the same. Therefore, any measured ID discrepancy between two estimates (e.g., for the same system modeled with two different force fields) is meaningful only if both were computed with the same $N$.

These results show that the sample-size dependence of TWO-NN can be used to estimate the geometric dimension of a dataset. The spread across truncation orders provides a practical measure of model uncertainty, while additional tests against fit range and resampling can further assess the robustness of the estimate when the geometric dimension is not already known.

\section*{Data Availability Statement}
The simulation data and analysis scripts supporting this study are available on Zenodo, at the URL https://doi.org/10.5281/zenodo.22282095.

\section*{Supplementary Material}
The supplementary material contains the complete parameters of the finite-size fits obtained at different truncation orders, the statistical bias of the TWO-NN regression estimator, the calculation of residence times, the details of the test with a noisy hyperplane, and the Ramachandran distributions sampled by the ten independent molecular dynamics trajectories.

\begin{acknowledgments}
The author thanks Toni Giorgino, Andrea Gardin, and GiovanniMaria Piccini for useful discussions, and Carlo Camilloni and Bruno Stegani for reading the manuscript. 
\end{acknowledgments}

\bibliography{biblio}

\end{document}

\title{Supplementary Materials for:\\
Derivation of the Sample-Size Scaling of TWO-NN Intrinsic-Dimension Estimates from Molecular Dynamics Trajectories}

\author{Riccardo Capelli}
\email{riccardo.capelli@unimi.it}
\affiliation{
Department of Biosciences,
Universit\`a degli Studi di Milano,
Via Celoria 26,
I-20133 Milan,
Italy
}

\date{\today}

\maketitle

\section{Finite-size fit parameters}

The finite-size data were fitted using the model introduced in the main text,
\begin{equation*}
    \mathrm{ID}_K(N)=\mathrm{ID}_{\infty}-\sum_{k=1}^{K}A_{2k}N^{-2k/\mathrm{ID}_{\infty}}.
\end{equation*}
Here, $A_{2k}=-C_{2k}$ relative to the notation used in the main text. Table~\ref{tab:id_infty_fits} reports the results obtained by treating $\mathrm{ID}_{\infty}$ and all correction coefficients as free parameters.

\begin{table}[h!]
\caption{Parameters obtained from the unconstrained fits of the finite-size model. The truncations $K=1$, $2$, and $3$ retain corrections through $r^2$, $r^4$, and $r^6$, respectively. Fits were weighted using the standard errors of the $A_N$ estimates defined in the main text. Reported parameter uncertainties are the asymptotic standard errors returned by the fit. Values of $\chi^2_{\mathrm{red}}$ describe the residuals only: estimates at different $N$ from the same replica are correlated, so they cannot be used as a goodness-of-fit test.}
\label{tab:id_infty_fits}
\small
\renewcommand{\arraystretch}{1.15}
\begin{tabular*}{\textwidth}{@{\extracolsep{\fill}}lccccc@{}}
\toprule
Truncation & $\mathrm{ID}_{\infty}$ & $A_2$ & $A_4$ & $A_6$ & $\chi^2_{\mathrm{red}}$ \\
\midrule
$K=1$ ($r^2$) & $24.81\pm0.02$ & $24.65\pm0.03$ & -- & -- & 23.0 \\
$K=2$ ($r^4$) & $23.47\pm0.09$ & $21.28\pm0.23$ & $3.12\pm0.22$ & -- & 1.28 \\
$K=3$ ($r^6$) & $25.30\pm0.74$ & $28.70\pm2.87$ & $-12.64\pm5.79$ & $11.38\pm4.01$ & 0.06 \\
\bottomrule
\end{tabular*}
\end{table}

\section{Statistical bias of the estimator}

To isolate the statistical bias of the regression estimator from geometric effects, we applied it to $\mu$ values drawn independently from the Pareto distribution $1-F(\mu)=\mu^{-d}$, for which the estimator would be exact in the limit of infinite $N$. Because $\log\mu$ is exponentially distributed with mean $1/d$, the relative bias is independent of $d$, as confirmed for $d=8$, 12, 16, and 24. With $F_i=i/N$, $i=0,\dots,m-1$, as in scikit-dimension 0.3.4, the relative bias for $d=24$ is $-1.45\%$, $-0.70\%$, $-0.31\%$, and $-0.15\%$ at $N=100$, 200, 500, and 1000, i.e., about $-1.45/N$; the values for $d=8$, 12, and 16 agree within the statistical uncertainty. With $F_i=i/N$, $i=1,\dots,m$, as in the current release, it is $+1.83/N$.

\section{Residence times}

Each configuration of the ten production trajectories (saved every 0.1~ns) was assigned to the $\phi<0$ region ($C7_{\mathrm{eq}}$ and C5) or to the $\phi>0$ region ($C7_{\mathrm{ax}}$) from the backbone dihedral angle $\phi$. To account for periodicity, values $\phi>150^\circ$ were shifted by $-360^\circ$, so that configurations of the C5 region near $\phi=\pm180^\circ$ are assigned to $\phi<0$; only 5 of the $10^7$ configurations have $95^\circ<\phi<165^\circ$. To avoid counting rapid recrossings of the barrier as transitions, a configuration changes state only when $\phi$ enters the core of the other state, $\phi<-30^\circ$ or $\phi>30^\circ$; between these values it keeps the previous assignment. A residence time is the duration of an uninterrupted stay in one state; the first and last stays of each trajectory, which are truncated, were discarded. The mean residence time is 1.75~$\mu$s in the $\phi<0$ region (median 1.26~$\mu$s, 541 stays) and 51~ns in $C7_{\mathrm{ax}}$ (median 36~ns). Each trajectory contains about 110 transitions, and $C7_{\mathrm{ax}}$ accounts for 2.8\% of the configurations. The results are unchanged for core boundaries between $\pm17^\circ$ and $\pm40^\circ$.

\section{Noisy hyperplane}

Points were drawn uniformly in a five-dimensional unit hypercube and embedded in $D=100$ features. Gaussian noise with standard deviation $\sigma$ was added to all features. For each $N$ between $10^2$ and $10^5$, between 3 and 60 independent samples were analyzed with the estimator used for the molecular data, including the bias correction. The fits assume a relative standard deviation of $1.2/\sqrt{N}$ for a single estimate, as measured on the molecular data. Because of the boundaries of the hypercube, the geometric correction was taken proportional to $N^{-1/d}$; the cross term proportional to $N^{+1/d}$ (Sec.~II of the main text) was neglected. The results are reported in Table~\ref{tab:noise} and in Fig.~4 of the main text.

\begin{table}[h!]
\caption{Five-dimensional hypercube embedded in $D=100$ features, with Gaussian noise of standard deviation $\sigma$ on each feature; the true dimension is 5. $\mathrm{ID}(10^2)$ and $\mathrm{ID}(10^5)$ are the bias-corrected estimates at the smallest and largest $N$. The other columns report fits of the estimates for $10^2\le N\le10^5$, either with the geometric correction only, $\mathrm{ID}(N)=d+C_1N^{-1/d}$, or with the geometric correction and the noise term of Eq.~(6), $\mathrm{ID}(N)=d+C_1N^{-1/d}+B_2N^{+2/d}$. In both functions, $d$ is the fitted dimension of the hyperplane, to be compared with the true value 5. For the hypercube, the geometric correction is proportional to $N^{-1/d}$ because of the boundaries (Sec.~II of the main text). $\chi^2_{\mathrm{red}}$ is the reduced chi-square of each fit: values much larger than 1 mean that the function does not describe the data. Uncertainties of $d$ are multiplied by $\sqrt{\chi^2_{\mathrm{red}}}$ when $\chi^2_{\mathrm{red}}>1$.}
\label{tab:noise}
\small
\renewcommand{\arraystretch}{1.15}
\begin{tabular*}{\textwidth}{@{\extracolsep{\fill}}ccccccc@{}}
\toprule
 & & & \multicolumn{2}{c}{Geometric correction only} & \multicolumn{2}{c}{Geometric correction and noise term} \\
\cmidrule(lr){4-5}\cmidrule(lr){6-7}
$\sigma$ & $\mathrm{ID}(10^2)$ & $\mathrm{ID}(10^5)$ & $d$ & $\chi^2_{\mathrm{red}}$ & $d$ & $\chi^2_{\mathrm{red}}$ \\
\midrule
0     & 4.23 & 4.85  & $5.04\pm0.03$ & 3.9 & $5.25\pm0.08$ & 1.8 \\
0.002 & 4.42 & 5.67  & $6.29\pm0.18$ & 57  & $4.98\pm0.04$ & 0.8 \\
0.005 & 4.62 & 9.54  & $40\pm22$     & 504 & $5.06\pm0.03$ & 1.0 \\
0.01  & 5.33 & 19.26 & diverges      & --  & $5.94\pm0.15$ & 18  \\
\bottomrule
\end{tabular*}
\end{table}

\clearpage
\begin{figure}[p]
  \includegraphics[width=\linewidth]{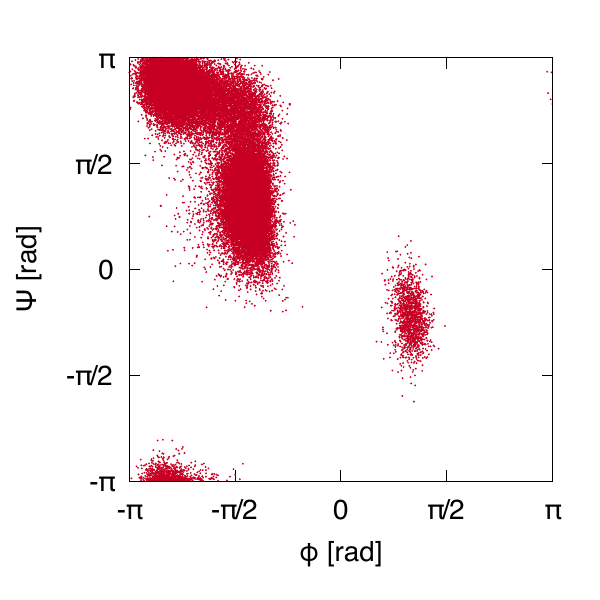}
  \caption{Ramachandran plot of the combined unbiased ensemble obtained from the ten independent $100~\mu$s trajectories of alanine dipeptide. Each point is a sampled configuration projected onto the backbone dihedral angles $\phi$ and $\psi$, with both angles wrapped to the interval $[-\pi,\pi]$. The populated region near $(\phi,\psi)=(-\pi/2,\pi/2)$ corresponds to the $C7_{\mathrm{eq}}$ basin, whereas the smaller region near $(\pi/3,-\pi/3)$ corresponds to the $C7_{\mathrm{ax}}$ basin. Points close to opposite plot boundaries are connected by the periodicity of the dihedral angles. This projection is shown only to document conformational sampling.}
  \label{fig:ramachandran}
\end{figure}